\documentclass{rspublic}
\usepackage{graphicx}

\begin{document}

\title[Stability analysis of three dimensional breather solitons...]
{Stability analysis of three dimensional breather solitons in a Bose--Einstein
Condensate}

\author[M. Matuszewski, E. Infeld, G. Rowlands, and M. Trippenbach]
{M. Matuszewski$^{1}$, E. Infeld$^{2}$, G. Rowlands$^{3}$, and M. Trippenbach$^{1,2}$}

\affiliation{$^{1}$ Institute of Theoretical Physics, Physics Department, Warsaw
University, Ho\.{z}a 69, PL-00-681 Warsaw, Poland, \\
$^{2}$ Soltan Institute for Nuclear Studies, Ho\.{z}a 69, PL-00-681 Warsaw, Poland. \\
$^{3}$ Department of Physics, University of Warwick, Coventry, CV4 7AL, UK}

\label{firstpage}

\maketitle

\begin{abstract}{Bose-Einstein condensate; breather solitons; optical lattice; Feshbach resonance}
We investigate the stability properties of breather soliton trains in a three-dimensional
Bose-Einstein Condensate with Feshbach Resonance Management of the scattering length.
This is done so as to generate both attractive and repulsive interaction. The condensate
is confined only by a one dimensional optical lattice and we consider both strong,
moderate, and weak confinement. By strong confinement we mean a situation in which a
quasi two dimensional soliton is created. Moderate confinement admits a fully three
dimensional soliton. Weak confinement allows individual solitons to interact. Stability
properties are investigated by several theoretical methods such as a variational
analysis, treatment of motion in effective potential wells, and collapse dynamics. Armed
with all the information forthcoming from these methods, we then undertake a numerical
calculation. Our theoretical predictions are fully confirmed, perhaps to a higher degree
than expected. We compare regions of stability in parameter space obtained from a fully
3D analysis with those from a quasi two-dimensional treatment, when the dynamics in one
direction are frozen. We find that in the 3D case the stability region splits into two
parts. However, as we tighten the confinement, one of the islands of stability moves
toward higher frequencies and the lower frequency region becomes more and more like that
for quasi 2D. We demonstrate these solutions in direct numerical simulations and,
importantly, suggest a way of creating robust 3D solitons in experiments in a Bose
Einstein Condensate in a one-dimensional lattice.

%%PACS numbers 03.75.-b, 03.75.Lm, 05.45.Yv, 42.65.Tg

\end{abstract}

\section{Introduction}

The creation of Bose--Einstein condensates (BEC) in vapours of alkali metals offers a
splendid opportunity to investigate the wide range of nonlinear effects invloving atomic
matter waves. A particular challenge is to develop methods for creating and controlling
matter-wave solitons. Both dark and bright one dimensional solitons in harmonic traps
have been observed (Burger \textit{et. al.} 1999; Denschlag \textit{et. al.} 2000;
Strecker \textit{et. al.} 2002; Khaykovich \textit{et. al.} 2002; Eiermann \textit{et.
al.} 2004). A promising approach to obtaining multidimensional solitons consists in
varying the scattering length (SL)\ of interatomic collisions. This can be achieved by
means of sweeping an external magnetic field through the zero-SL point close to the
Feshbach resonance (Donley \textit{et. al.} 2001; Inouye \textit{et. al.} 1998; Fedichev
\textit{et. al.} 1996; Saito \& Ueda 2002; Theis \textit{et. al.} 2004). The application
of an ac magnetic field may induce a periodic modulation of the SL, opening the way to so
called `Feshbach-resonance management' (FRM) (Kevrekidis \textit{et. al.} 2003). A
noteworthy FRM-induced effect is the possibility of creating self-trapped oscillating BEC
solitons (breathers) without an external trap in the 2D case.  The underlying idea is
that fast modulations create an effective potential on a slower timescale. This potential
can stabilize the soliton.

The mathematical model for a BEC based on the Gross-Pitaevskii equation (GPE)
with harmonic modulation of
the SL was investigated by Saito \& Ueda (2002, 2004), Abdullaev \textit{et. al.} (2003), and
Montesinos \textit{et. al.} (2004).
The conclusion was that FRM enables us
to stabilize 2D breather solitons even \textit{without} the use of an external trap. According to
these references, 3D breathers conversely require at least a tight, one dimensional harmonic trap,
practically reducing the problem to 2D (G\"orlitz \textit{et. al.} 2001).
Later on we will call this approach a
quasi two dimensional treatment (Q2D). It will be defined more precisely below.
Stabilization of
3D droplets is also possible by including dissipative effects (Saito \& Ueda 2004).

%We show that the relation of the 3D stability region in the parameter space to that of 2D
%treatment is at first surprising, as they seem to be to some extent unrelated and that of 3D is
%not a subset of the 2D stability region. In systems with FRM, an additional degree of freedom can
%stabilize, contrary to most known physical situations. {\it Therefore in general one should be
%wary of using 2D treatment to infer the 3D stability properties.}

Trippenbach \textit{et. al.} (2004) demonstrated that 3D solitons can be stabilized by a
combination of FRM and a 1D optical lattice (1D OL), instead of a 1D harmonic trap. This
issue has practical relevance, as a 1D OL can easily be created by illuminating the BEC
by a pair of counterpropagating laser beams so that they form a periodic interference
pattern (Stecher \textit{et. al.} 1997). Incidently, it is easier to realize a tight
confinement configuration in an optical lattice than in a harmonic trap. Hence this
environment may well be more friendly for creating quasi 2D solitons. The lattice will be
weak or strong depending on the intensity of the laser light. The combined OL-FRM
stabilization of 3D solitons is possible even in a weak lattice, when atoms confined in
different cells do interact.

By analyzing the
stability charts in configuration space we find two distinct regions where stable solutions
exist. The first of these regions has its counterpart in the Q2D treatment. The other region
appears when the frequency of modulation exceeds the lowest excitation frequency of the confining
potential. It is not present in the Q2D treatment; and so can only correspond to fully 3D
solitons. In the limit of tight confinement, the latter region moves up to extremely high
frequencies and the Q2D stability chart is recovered.

\section{Theoretical approach}

\begin{figure}[tbp]
\includegraphics[width=8.5cm]{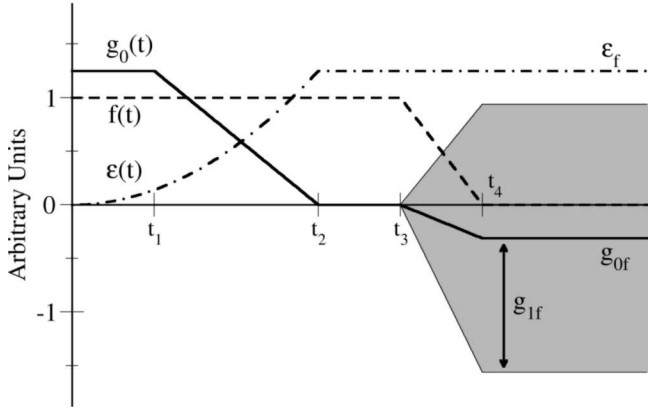}
\caption{The time dependence of the nonlinear coefficient, $g$, switching-off function, $f(t)$,
and optical-lattice strength, $\protect\varepsilon ,$ in the numerical experiment. This
experiment leads to
the establishment of stable 3D breathing solitons supported by the combination of the quasi-1D
lattice and Feshbach-resonance management (FRM). The shaded area indicates rapid oscillations of
$g(t)$, which account for the FRM.} \label{fig1}
\end{figure}

We describe our system by the GPE in reduced units, including a time-dependent (FRM-controlled)
nonlinear coefficient $g(t)$ and an external potential $U(\mathbf{r},t)$
\begin{equation}
\ri\psi _{t}=\left[ -(1/2)\nabla ^{2}+U(\mathbf{r},t)+g(t)|\psi |^{2}\right] \psi .  \label{NLS}
\end{equation} Initially the BEC is in the ground state of a radial (2D)
parabolic trap with frequency $\omega _{\perp }$, supplemented, in the longitudinal direction, by
`end caps', induced by transverse light sheets. The configuration is very similar to
that used to
create soliton trains in a Li${^{7}}$ condensate (Khaykovich \textit{et. al.} 2002).
A 1D lattice potential in the
axial direction is adiabatically turned on from $\varepsilon=0$
to $\varepsilon=\varepsilon _{f}$, see figure~\ref{fig1}. Thus, the full potential is,
with period normalized to $\pi$
\begin{equation}
U(\mathbf{r},t)=\varepsilon (t)\left[ 1-\cos (2z)\right] +f(t)\left[(1/2)\omega _{\perp
}^{2}\varrho ^{2}+U_{0}(z)\right],  \label{V}
\end{equation}
where $\varrho $ is the radial variable in the plane transverse to $z$, and the axial
`end-cap' potential, $U_{0}(z)$ is approximated by a sufficiently deep one dimensional
rectangular potential well. The width of the well determines the number of peaks in the
final structure. The $f(t)$ is a switch function (see figure~\ref{fig1}).

The nonlinear interaction coupling is described by
\begin{equation}
g(t)=g_{0}(t)+g_{1}(t)\sin (\Omega t).  \label{g}
\end{equation}
Initially $g_1(0)=0$ and $g(0)=g_0(0)>0$. At some moment $t_{1}$, we
begin decreasing $g_0(t)$ linearly. It vanishes at time $t_{2}$, and
remains zero up to $t_{3}$, when we start to gradually switch on the
rapid FRM modulation of the SL. In the interval $[t_3,t_4]$,
$g_0(t)$ decreases linearly from zero to a negative value $g_{0f}$
and the amplitude of the modulation $g_1(t)$ increases from zero to
$g_{1f}$, see figure~\ref{fig1}. Simultaneously,  both the radial
confinement and end-caps are gradually switched off (see the
behaviour of the $f(t)$ function in figure~\ref{fig1}). At times
$t>t_{4}$, $g(t)$ oscillates with a constant amplitude $g_{1f}$
around a negative average value $g_{0f}$. Consequently, a soliton so
created, if any, is supported by the combination of the 1D lattice
and FRM. Note that we choose a set of specific ramp functions $g_0$
and $g_1$ and they grow rather rapidly in time.  In general the
stability diagram may depend on the shape and rapidity of the ramp
functions. We did not investigate this issue in detail, but we
believe that our results are to some extend universal. The reason we
start from a positive value of $g_0$ is so that then the atoms can
be uniformly distributed in the cells. This will allow stabilization
of solitons in all filled cells simultaneously.

Numerical experiments following the path outlined in figure~\ref{fig1} indicate that it is possible
to create stable 3D solitons (see inset to figure~\ref{fig2}). Before displaying the
results, we first resort to the variational approximation (VA) in order to predict conditions on
the modulation frequency and the size of the negative average nonlinear coefficient $g_{0f}$,
necessary to support 3D solitons.

\section{Variational analysis}

The VA can be applied to the description of BEC dynamics under diverse circumstances (e.
g. Saito \& Ueda 2003; Montesinos \textit{et. al.} 2004; Baizakov  \textit{et. al.}
2003). Equation~(\ref{NLS}) is derived from the Lagrangian density
\begin{eqnarray}
{\cal{L}}= i(\psi _{t}^{\ast }\psi -\psi _{t}^{\ast }\psi )-|\psi _{\varrho }|^{2}-|\psi _{z}|^{2}
-g(t)|\psi |^{4} - 2U|\psi |^{2} \label{L}
\end{eqnarray}
We use the VA for $t>t_3$ and choose a hybrid ansatz composed of an hyperbolic secans
function and a Gaussian for the solution for one lattice cell (calculations with two
Gaussians give slightly inferior results, in terms of agreement with numerics). The
amplitude is $A(t)$, the overall phase is $\phi$, the radial and axial widths are $W(t)\
$ and $V(t)$ respectively, and $b(t)$ and $\beta (t)$ are the corresponding chirps
\begin{equation}
\psi (\mathbf{r},t)=A \mathrm{sech}(\varrho/W)
\re^{\left[ -\ri b\varrho^{2} -( 1/2V^{2}+\ri\beta)z^{2}+\ri\phi\right]}.
\label{ansatz}
\end{equation}
The reduced Lagrangian can be found upon substituting~(\ref{ansatz}) into~(\ref{L}) and
integrating over space. It is
\begin{eqnarray}
&L & = A^2W^2V \, \bigg[ \frac{ A^2 g I_3}{2\sqrt{2}}+\frac{\varepsilon I_1}{ e^{V^2}} -
\frac{f(t)\omega_\perp^2 W^2 I_2}{2} -\frac{I_1}{4 V^2} +  \nonumber  \\ & - & \frac{V^2 I_1}{4}
\left(\beta^2 + \dot{\beta}\right) -\frac{I_1 - I_3}{2 W^2}
-\frac{W^2 I_2}{2}\left(b^2+\dot{b} \right) - I_1 \dot{\phi} \bigg],
\end{eqnarray}
where $I_1=2 \pi \ln 2$, $I_2=(9/4) \pi \zeta(3)$, and $I_3=(\pi/3)(4 \ln 2 - 1)$.
By varying this reduced Lagrangian with respect to $\phi$ we obtain the constant $E=A^{2}W^{2}V
=(2\pi ^{3/2} \ln 2)^{-1}\int_{cell}|\psi |^{2}\rd\mathbf{r}=(2 n \pi ^{3/2} \ln 2)^{-1}$,
where the integral extends over one
cell of the lattice, and $n$ is the number of occupied lattice cells. Notice that the total
number of
atoms is included in the definition of the nonlinear coupling $g(t)$, and the total wavefunction is
normalized to unity. When the other four variational equations are derived, we can deduce two
dynamical equations for the widths by substituting for the chirps
\begin{eqnarray}
b&=& \dot{W}/W, \\
\beta&=& \dot{V}/V,
%% \ddot{W} &=&\frac{1}{W^{3}}-f(t)\omega _{\perp }^{2} W+ \frac{Eg(t)}{\sqrt{8}W^{3}V},
%% \label{variat1} \\
%% \ddot{V} &=&\frac{1}{V^{3}}-4\varepsilon_f V\exp \left( -V^{2}\right)
%% +\frac{Eg(t)}{\sqrt{8}W^{2}V^{2}}.  \label{variat2}
\end{eqnarray}
to obtain
\begin{eqnarray}
\ddot{W} &=&\frac{J_1}{W^{3}}-f(t)\omega _{\perp }^{2} W+ \frac{J_2 g(t)}{W^{3}V},
\label{variat1} \\
\ddot{V} &=&\frac{1}{V^{3}}-4\varepsilon_f V\exp \left( -V^{2}\right)
+\frac{J_3 g(t)}{W^{2}V^{2}},  \label{variat2}
\end{eqnarray}
%% where $J_1=(8 \ln 2 + 4) / (27 \zeta(3))$, $J_2=\sqrt{2}(4 \ln 2 -1) / (27 n \ln 2 \zeta(3))$,
%% and $J_3=\sqrt{2}(4 \ln 2 -1) / (24 n \ln^2 2)$.
where $J_1=(I_1-I_3)/I_2$, $J_2=E I_3/(\sqrt{2} I_2)$, and $J_3=E I_3/ (\sqrt{2} I_1)$.
All $J$'s are positive. These equations describe the dynamics of a single peak, and so
with small corrections can be applied to the problem of a BEC confined in a 1D harmonic
trap (Saito \& Ueda 2003; Abdullaev \textit{et. al.} 2003; Montesinos \textit{et. al.}
2004). Note that for $f(t)=0$ one can derive a simple condition: $J_2 V\ddot{V} -J_3
W\ddot{W} =J_2 V^{-2}-J_1 J_3 W^{-2}+4\varepsilon V^2 \exp(-V^2)$. This equation depends
on time only implicitly, and is used when describing collapse scenarios
(Section~\ref{ap1}).

In the corresponding Q2D treatment we drop the $z$ dimension in equation~(\ref{NLS}). We
assume that in this direction the profile of the wavefunction is fixed and reproduces the
ground state $\psi_0$ of the single lattice cell or harmonic potential, as in Saito \&
Ueda (2003) and Montesinos \textit{et. al.} (2004). The reduced potential in 2D GP will
take the form $U(\varrho,t)= f(t)(1/2)\omega _{\perp }^{2}\varrho ^{2}$. In the VA we
take $V$ constant, $V_0$ as found from the equation $4\varepsilon_f V_{0}^{4}\exp \left(
-V_{0}^{2}\right)=1$, and only solve equation~(\ref{variat1}). In numerical simulations
we rescale the nonlinear coupling coefficient $g_{\mathrm{2D}} = g \times (\int
|\psi_0|^2 \psi_0 \rd z)/(\int \psi_0 \rd z)$.

\section{Numerical results}

We simulated both the full GPE, equation~(\ref{NLS}), using an axisymmetric code (for 3D), a Cartesian
code (for Q2D), and the variational equations for comparison. Numerical simulations followed the
path outlined in figure~\ref{fig1}.

\begin{figure}[tbp]
\includegraphics[width=8.5cm]{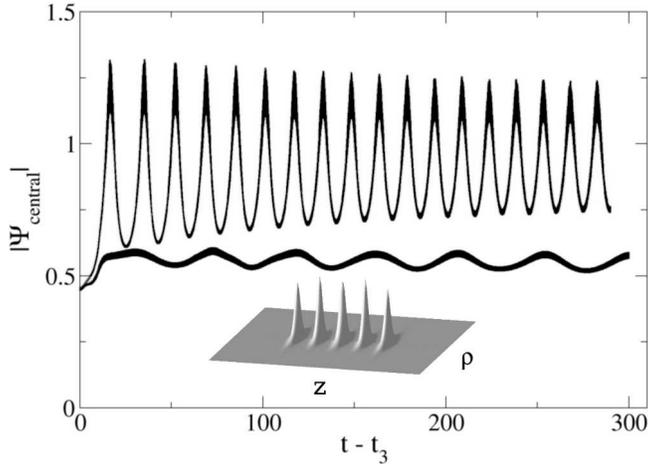}
\caption{Comparison of the central peak amplitude evolution in the
3D treatment (lower curve) and the corresponding Q2D treatment
(upper curve). This is an example of a moderate confinement, when
only a fully 3D treatment describes the dynamics of the system
properly. The normalized parameters are $g_{0 \mathrm{f}}=-22$,
$g_{1 \mathrm{f}}=4g_{0 \mathrm{f}}$, $\varepsilon_{\mathrm{f}}=50$,
$\Omega =36$, $\protect\omega _{\perp }=0.3$, $t_{1}=30$,
$t_{2}=100$, $t_{3}=120$, and $t_{4}=130$. The inset shows the
structure in the 3D treatment (in the Q2D treatment $z$ dimension is
dropped). The unit of time is $m\protect\lambda ^{2}/(\protect\pi
^{2}\hbar $). Stabilization in the $z$ direction appears to be by
attractive interaction. The soliton in this direction is discrete
rather than gap-type.} \label{fig2}
\end{figure}

An example of the numerical results in the moderate confinement
regime is shown in figure~\ref{fig2}. The parameters used in the
simulations would correspond, for $^{87}$Rb atoms, to an OL period
of $\lambda =1.5\3\mu $m, an initial radial-confinement frequency of
$\omega _{\perp }=2 \pi \times 160\3$Hz, an FRM frequency of
$\Omega=2 \pi \times 19\3$kHz, a lattice depth of $\varepsilon_f
=25$ recoil energies, an effective nonlinear coefficient of $Na=\pm
\,2\cdot 10^{-5}\3$m, where $N$ is the number of atoms per lattice
cell, with a total number of atoms in the range of $10^{4}-10^{6}$.
The corresponding values of the normalized parameters are given in
the figure captions. This figure shows the evolution of the
central-peak's amplitude versus time in dimensionless units, defined
by equation~(\ref{NLS}). After an initial transient, a stable
breathing structure is established. The difference between the
dynamics in the 3D treatment (lower curve) and the corresponding Q2D
treatment (upper curve) is obvious. This is an example of a moderate
confinement, when only the 3D treatment describes the dynamics of
the system properly. The inset shows the 3D soliton structure for a
fixed moment of time.

\begin{figure}[tbp]
\includegraphics[width=8.5cm]{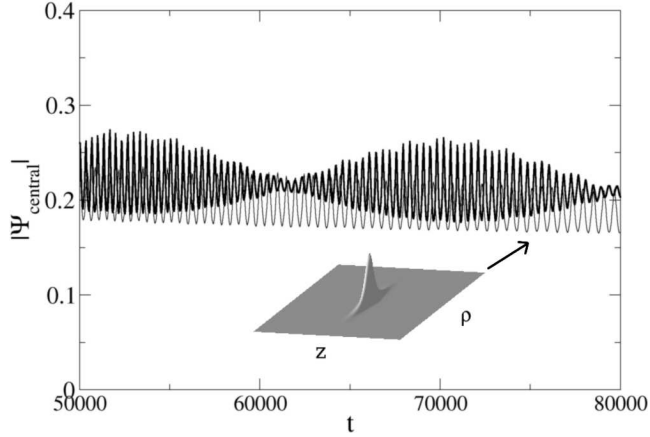}
\caption{The thick and thin lines show the
evolution of the amplitude of the central peak, respectively, of the three-peaked soliton, and in
the case when, at $t=7500$, the side peaks were suddenly removed (the latter configuration is
shown in the inset). This is an example of a weak confinement, when individual solitons
do interact.
The normalized parameters are $g_{0\mathrm{f}}=-18$, $g_{1\mathrm{f}}=4g_{0\mathrm{f}}$,
$\protect\epsilon _{\mathrm{f}}=20.5$, $\Omega =22$, $\protect\omega _{\perp }=0.3$, $t_{1}=30$,
$t_{2}=100$, $t_{3}=120$, and $t_{4}=130$. }
\label{interaction}
\end{figure}

The dynamics in the weak confinement regime are displayed in figure~\ref{interaction}.
The figure reveals the influence of neighbouring solitons on the amplitude of the central
peak. The thick curve corresponds to the evolution of the full multipeak structure,
similar to that shown in the inset of figure~\ref{fig2}. The interaction between
neighbouring solitons can be seen in the form of a beat. This phenomenon can be explained
by the fact that the oscillation periods of the adjacent solitons are slightly different
as there is a small difference in the number of atoms in neighbouring cells. To obtain
the thin curve we repeated the above calculations up to the time $t=7500$ and then
removed all but the central peak. The structure so formed can be seen the inset of
figure~\ref{interaction}. The beat is no longer visible. The difference between these two
cases proves that the dynamics of stable breathers as described here is a collective
multi-peak phenomenon.

\begin{figure}[tbp]
\includegraphics[width=12cm]{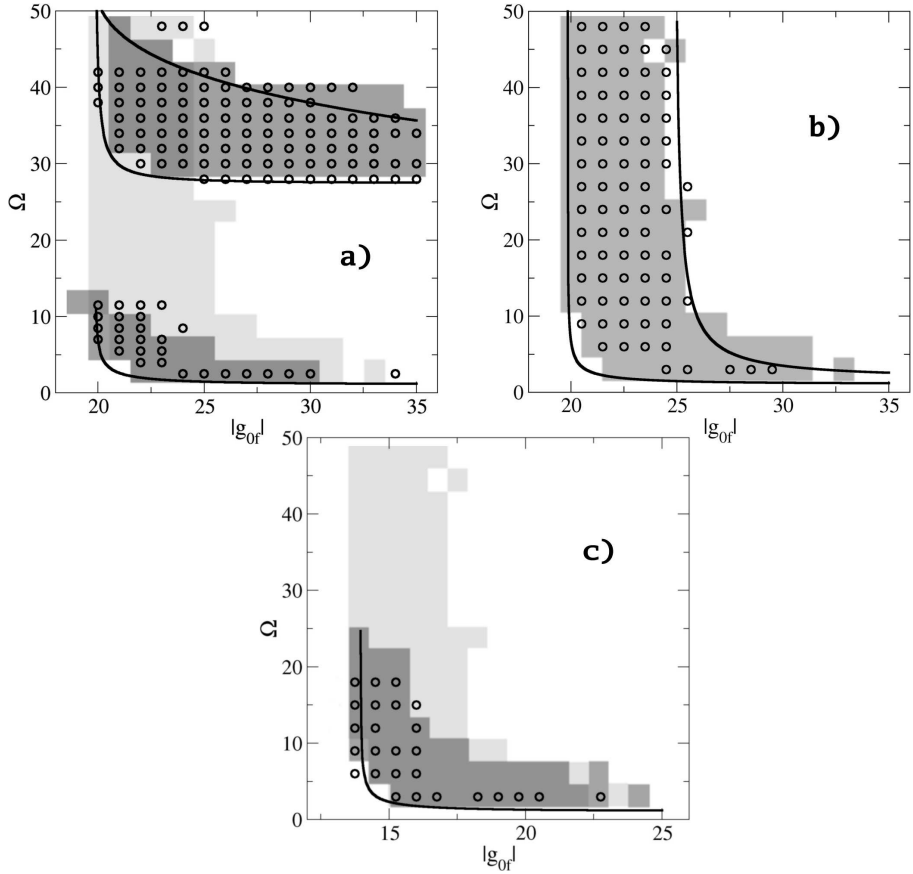}
\caption{Stability regions for solitons in the $\left( |g_{0f}|,\Omega \right)$ plane, as
predicted by the variational approximation (shaded area), and found from direct
simulations of the Gross-Pitaevskii equation (circles). Both a) corresponding to
$\epsilon _{\mathrm{f}}=50$ and $\Omega_0=26.76$ and c) corresponding to $\epsilon
_{\mathrm{f}}=200$ and $\Omega_0=55.06$ were obtained from a 3D analysis and b) is the result
of a Q2D treatment. The other parameters are as in figure~\protect\ref{fig2}. The lower
regions in a) and c) correspond to Q2D solitons and those corresponding to the upper
regions are fully 3D. As we see, as the confinement increases, the lower region obtained
from the 3D analysis becomes more and more like that following from Q2D. The curves
limiting stability regions were obtained analytically, see equations~(\ref{con1}),
(\ref{xiscal}), (\ref{con3}), and~(\ref{twoomega}). The light-gray area in a) and c) is the stability
region in the Q2D treatment appropriately rescaled. } \label{fig3a}
\end{figure}

%In the 3D treatment we used both Eqs.~(\ref{variat1}) and~(\ref{variat2}).
In figure~\ref{fig3a} we collected results of a systematic scan of parameter space based
on GPE simulations and compared them with the predictions of the VA (a similar analysis
can be performed if we replace 1D OL with a 1D harmonic trap -- the conclusion does not
depend on the form of confinement in the $z$ direction). The agreement between VA and
direct simulations is very good. The borders of the VA stability regions were found
analytically e.~g.~ upon considering effective potentials. A detailed derivation is
presented in \S\ref{stan}. As seen from figure~\ref{fig3a} a), in a fully 3D treatment we
have found \textit{two islands of stability}. Note the similarity of the lower regions of
a) and c) to that of figure~\protect\ref{fig3a} b), which portrays the results of the Q2D
treatment. This region corresponds to Q2D solitons. On the other hand, the upper region
contains fully 3D solitons. They appear when the frequency of modulation exceeds the
lowest excitation frequency of the confining potential. As we see in figure~\ref{fig3a},
when the strength of the lattice $\varepsilon_f$ is increased, this region moves towards
higher frequencies (and has risen beyond the scope of c)), the Q2D region expands
upwards, and the whole picture becomes more and more like figure~\protect\ref{fig3a} b).

%\begin{figure}[tbp]
%\includegraphics[width=8.5cm]{ppsd2.eps}
%\caption{Same as figure~\protect\ref{fig3a}, but for Q2D treatment. The curve on the left corresponds
%to equation~(\ref{con1}), and the right to equation~(\ref{xiscal}).} \label{fig3b}
%\end{figure}
%
%\begin{figure}[tbp]
%\includegraphics[width=8.5cm]{ppsd3ab.eps}
%\caption{Same as figure~\protect\ref{fig3a}, but for $\epsilon _{\mathrm{f}}=200$.} \label{fig4}
%\end{figure}

\section{Stability analysis} \label{stan}

We base our calculations on the variational equations~(\ref{variat1}), (\ref{variat2})
with $f(t)=0$.
%\begin{eqnarray} \label{avariat1}
%  \ddot{W} &=& \frac{J_1}{W^{3}}
%    + \frac{J_2 \left(g_0+g_1 \cos(\Omega t)\right)}{W^3 V}, \\
%  \ddot{V} &=& \frac{1}{V^{3}} - 4 \varepsilon V e^{-V^2}
%  + \frac{J_3 \left(g_0 + g_1 \cos(\Omega t)\right)}{V^2 W^2}.\label{avariat2}
%%%   \ddot{W} &=&\frac{J_1}{W^{3}}-f(t)\omega _{\perp }^{2} W+ \frac{J_2 g(t)}{W^{3}V},
%%%   \label{variat1} \\
%%%   \ddot{V} &=&\frac{1}{V^{3}}-4\varepsilon_f V\exp \left( -V^{2}\right)
%%%   +\frac{J_3 g(t)}{W^{2}V^{2}},  \label{variat2}
%\end{eqnarray}
In the context of these equations, we will derive approximate formulas for borders of
stability regions in parameter space. Several types of instability will be considered.
Some of them can only be described by the full 3D treatment. All stability conditions
derived below lead to limiting curves in figure~\ref{fig3a}, with surprising accuracy.
They are in good agreement with the results of numerical simulations of both variational
equations~(\ref{variat1}), (\ref{variat2}) and the GPE~(\ref{NLS}).

\textit{Q2D treatment.} In the Q2D treatment we neglect the dynamics of $V$, assuming
$V=V_0$ and only consider the variational  equation for $W$, (\ref{variat1})
\begin{eqnarray} \label{var2D}
\ddot{W} &=&
%\frac{1}{W^{3}}\left[J_1 + \frac{J_2}{V_0}\left(g_0+g_1 \cos(\Omega t)\right)\right] =
\frac{1}{W^{3}}\left[-A + B \cos(\Omega t) \right].
\end{eqnarray}
Here we introduced the constants $A=J_1(g_0/g_c-1)$ and $B=-J_1 g_1/g_c$. The critical
nonlinearity $g_c=-V_0 J_1/J_2$ is equal to the threshold for collapse in the absence of
rapid modulations.

In the adiabatic approximation we separate $W$ into slowly and rapidly oscillating parts
$W=w+\delta$, and assume that the latter is relatively small, $\delta \ll w$.
Equation~(\ref{var2D}) takes the form
\begin{equation} \label{separ}
\ddot{w} + \ddot{\delta} = \left(\frac{1}{w^{3}}-\frac{3 \delta}{w^{4}}\right)
\left[-A + B \cos(\Omega t) \right].
\end{equation}
In the adiabatic limit of small $\xi=|\dot{w}|/(w \Omega)$ we obtain approximately
\begin{eqnarray}
\delta &=& -\frac{B}{w^{3} \Omega^2} \cos(\Omega t), \nonumber \\
%%\ddot{w} &=& - \frac{A}{w^{3}} - \frac{B}{w^{4}} \overline{\delta \cos(\Omega t)}.
\ddot{w} &=& - \frac{A}{w^{3}} + \frac{3 B^2}{2 w^7 \Omega^2}. \label{adiab}
\end{eqnarray}
The second formula is obtained using the first and can be interpreted as an equation of
motion of a particle in an effective potential
\begin{equation} \label{uef}
U_{\mathrm{ef}} = - \frac{C}{w^{2}} + \frac{D}{w^6},
\end{equation}
where $C=J_1\left(g_0/g_c-1\right)/2$ and $D=\left[J_1 g_1/(2 g_c \Omega)\right]^2$.

We denote the initial width $w_0$ and choose $\dot{w}_0=0$. From~(\ref{uef}) we see that
the particle will be trapped by the potential provided that $U_{\mathrm{ef}}(w_0)<0$, as
$U_{\mathrm{ef}}(\infty)=0$. This gives a stability condition
\begin{equation} \label{con1}
w_0 > \left(\frac{D}{C}\right)^{1/4},
\end{equation}
which leads to the lower limiting curve in figure~\ref{fig3a} b). Note that the above
requires $C>0$, or equivalently $g_0/g_c>1$. This was the {\it only} stability condition
known so far (Saito \& Ueda 2003; Montesinos \textit{et. al.} 2004).

We now consider if the adiabatic assumption used when deriving equations~(\ref{adiab}) is
consistent with the result, i.e. whether $\xi$, defined above, remains small. It's
maximum value can easily be estimated. If the oscillations in the potential~(\ref{uef})
are small, then  $|\dot{w}|$ is small and so is $\xi$. If on the other hand these
oscillations are large, one cycle can be split into two stages, (a) motion in the
potential $U_{\mathrm{ef}} \approx - C/w^{2}$ for large $w$, and (b) motion in the
potential $U_{\mathrm{ef}} \approx D/w^6$ for small $w$.  The $\xi$ coefficient reaches
its maximum value during the (b) stage. All possible trajectories in the (b) stage
potential are approximately described by a function, which scales with $D$ and the
initial conditions in the following way
\begin{equation} \label{xiscal}
\xi_{max} \sim \sqrt{D} \Omega^{-1} w_{min}^{-4}(w_0),
\end{equation}
where $w_{min}$ is the value of $w$ at the smaller turning point and can be calculated
from $w_0$, using equation (\ref{uef}). It turns out that if the right-hand side
of~(\ref{xiscal}) exceeds a critical value, an instability occurs in the numerical
simulations. We observe an increase in the particle energy (equation (\ref{adiab}) is no
longer valid) around the smaller turning points, and consequently particle escape from
the potential well~(\ref{uef}). We stress that, for $\xi_{max}$ less than critical, the
system is stable for a very long time. Thus we obtain the upper limit in
figure~\ref{fig3a} b).

\textit{3D treatment.} Now we extend our considerations to the case when the axial width
of the peak, $V$ is no longer constant and this influences stabilization. We keep the
assumption that $V$ is close to the width of the localized Wannier function of our
lattice potential.  Thus we write $V= V_0 + \eta$, where $\eta \ll V_0$.

The value of $V_0$ is taken such that the first two terms on the right hand side of
equation~(\ref{variat2}) cancel (there are in fact two such values and we take the lower
one, known to be stable). From the third, nonlinear term we take only the oscillating
part (the first two non-oscillating linear terms are assumed to dominate). In first order
in $\eta/V_0$ we end up with a harmonic oscillator type formula with a small driving term
\begin{equation} \label{dynv}
  \ddot{V} = \ddot{\eta} \approx -\Omega_0^2 \eta + \gamma \cos(\Omega t),
\end{equation}
where $\Omega_0^2 = 3/V_0^4 + 4 \varepsilon (1-2V_0^2) \exp(-V_0^2) = (2/V_0^2)^2 - 2/ V_0^2$
and $\gamma = J_3 g_1/(V_0^2 W^2)$. The solution is
\begin{equation} \label{eta}
  \eta = \frac{\gamma}{\Omega_0^2-\Omega^2} \cos (\Omega t).
\end{equation}
Now we rewrite equation~(\ref{variat1}), neglecting terms of higher order in $\eta/V_0$
\begin{equation}
\ddot{W} =\frac{J_1}{W^{3}}
  + \frac{J_2\left(g_0+g_1 \cos(\Omega t)\right)}{W^3 V_0}\left(1-\frac{\eta}{V_0}
  \right),
\end{equation}
and substitute $\eta$ from~(\ref{eta}). In the adiabatic limit, we repeat the reasoning
that led to equation (\ref{uef}), to obtain an equation of motion for a particle in an
effective potential, including a new term
\begin{equation} \label{uef2}
U_{\mathrm{ef}} = - \frac{C}{w^{2}} + \frac{F}{w^4} + \frac{D}{w^6},
\end{equation}
where $F=J_2 J_3 g_1^2/\left[8 V_0^4 (\Omega^2 - \Omega_0^2)\right]$. If the last term
in~(\ref{uef2}) can be neglected as compared to the middle one (this is true if $\Omega$
is close to $\Omega_0$), another stability condition can be found
\begin{equation} \label{con3}
w_0 > \left(\frac{F}{C}\right)^{1/2}.
\end{equation}
This gives the lower limit of the upper region in figure~\ref{fig3a} a).  Strictly
speaking, all terms in~(\ref{uef2}) should be taken into account when calculating these
stability conditions. However, approximate formulas~(\ref{con1}) and~(\ref{con3}) prove
to be surprisingly accurate for the range of parameters studied here. A condition
analogous to (\ref{xiscal}) will follow:
\begin{equation} \label{xiscal2}
\xi_{max} \sim \sqrt{F} \Omega^{-1} w_{min}^{-3}(w_0),
\end{equation}
However, another, lower border of the stability region can be found when considering
resonance with double frequency of the forcing term. In this case we have to include
higher, anharmonic terms in equation~(\ref{dynv}). In the following simple reasoning we
will just take the first, $\eta^2$ term,
\begin{eqnarray} \label{eq24}
\ddot{\eta} & \approx & -\Omega_0^2 \eta +\alpha \eta^2 + \gamma \cos(\Omega t),
\end{eqnarray}
where $\alpha  =  8 V_0^{-5} - 2V_0^{-3}$.  We define $\Omega=2 \Omega_0-\epsilon$. The
linear solution of \ref{eq24} is
\begin{equation}
  \eta^{(0)} = - \frac{\gamma}{3 \Omega_0^2} \cos[(2\Omega_0 - \epsilon)t],
\end{equation}
and the next order correction $\eta^{(1)}$, obtained when the quadratic term is included,
satifies
\begin{equation}
  \ddot{\eta}^{(1)} + \Omega_0^2\eta^{(1)} =
    2 \alpha \eta^{(0)} \eta^{(1)}.
\end{equation}
For a solution in the form $\eta^{(1)}=b \cos[(\Omega_0 -\epsilon/2)t]$ we get
\begin{equation}
  -(\Omega_0-\epsilon/2)^2 + \Omega_0^2 = -\frac{\alpha \gamma}{3 \Omega_0^2}.
\end{equation}
Thus the threshold for appearance of a strong resonance when the driving frequency $\Omega$
comes close to $2 \Omega_0$ is given by
\begin{equation} \label{twoomega}
  \epsilon = -\frac{\alpha \gamma}{3 \Omega_0^3}>0,
\end{equation}
where we take $\gamma = J_3 g_1/(V_0^2 w_{min}^2)$ and the minimal value of the width
during the evolution as $w_{min}=\sqrt[4]{D/C}$. In fact $w_{min}$ cannot be any smaller,
as it would imply another instability (compare to equation~\ref{con1}). A quadratic in
$\Omega$ is obtained. Simple manipulations show that $\epsilon$ tends to zero as $C$
tends to zero when $g_{0f}=g_{0c}\approx 20$ and $\Omega= 2 \Omega_0$ on figure
\ref{fig3a} a).

A fuller analysis would include an $\eta^3$ term in~(\ref{eq24}) and the result would
depend on $b$, the amplitude of $\eta^{(1)}$. This is why our $\epsilon$~(\ref{twoomega})
is just a threshold value. For a full analysis, see Landau \& Lifshitz (1960). Thus we
obtain the upper limiting curve figure~\ref{fig3a} a). The $\eta^3$ term indicates that
the resonance in question appears {\it above} this limiting curve in figure ~\ref{fig3a}
a).

Interestingly, the instability discussed in the Q2D analysis, associated with a breakdown
of the adiabatic approximation~(\ref{xiscal}), is strongly suppressed in the 3D treament
in the range $\Omega_0<\Omega<2\Omega_0$ (see equation \ref{xiscal2}). This is due to
flattening of the effective potential~(\ref{uef2}) by the additional term, and a decrease
of the value of the $\xi$ coefficient during the evolution.

\section{Collapse and spreadout} \label{ap1}

When considering possible collapse scenarios we look at the leading terms in
equation~(\ref{variat1}), when $\tau=t_{col}-t$ tends to zero. The $\sin(\Omega t)$ term
either tends to a constant or else to $\pm \Omega \tau$, in which latter case it can be
ignored in our considerations. We consider the leading small $\tau$ terms of
equation~(\ref{variat1}) to be
\begin{equation}
\ddot{W}=\frac{(J_1-\Gamma/V)}{W^3}, \,\,\,\,
\Gamma=-J_2\left(g_0+g_1 \sin(\Omega t_{col})\right)
 \label{A1}
\end{equation}
and of equation~(\ref{variat2})
\begin{equation}
\ddot{V}=\frac{1}{V^3}-\frac{\Gamma'}{(WV)^2}, \,\,\,\,
\Gamma'=-J_3\left(g_0+g_1 \sin(\Omega t_{col})\right)
. \label{A2}
\end{equation}
Now $\Gamma$ and $\Gamma'$ can take either common sign, but we will see that only
$\Gamma,\Gamma' > 0$ can lead to collapse. As $\tau \rightarrow 0$ either all three terms
in equation~(\ref{A1}) cancel, or else two predominate and cancel in lowest order $\tau$.
When all three cancel, $W \sim \tau^{1/2}$, $V \rightarrow V_{col}$ (1). When two terms
cancel, either the first and third are of lowest order and $W \sim V \sim \tau^{2/5}$
(2), or else the second and third cancel because $V_{col}=\Gamma/J_1$ and we find that $W
\sim V- V_{col} \sim \tau^{2/3}$ (3). We now look at all three possibilities in some
detail.

(1) All three terms in equation~(\ref{A1}) are of equal strength. We find
from both equations~(\ref{A1}) and~(\ref{A2})
\begin{eqnarray}
W & = & \alpha \tau^{1/2} + \frac{\beta \Gamma}{4 \alpha^3 V_{col}^2} \tau^{3/2}
\left[(\ln \tau)^2 - 3 \ln \tau\right],
%%+ \frac{\beta \Gamma}{4\alpha^2 V_{col}^2}
%%\tau^{3/2}\left[
%%(\ln\tau)^2-3 \, \ln\tau\right] \label{A3}
\\
{V} & = & V_{col}+\beta (\tau \ln\tau-\tau), \label{A4}
\end{eqnarray}
where $\alpha=\sqrt{2}(\Gamma/V_{col}-J_1)^{1/4}$, $\beta = \Gamma'/(\alpha^2
V_{col}^2)$, $\Gamma>V_{col}J_1$.

(2) When the first and the third term in equation~(\ref{A1}) cancel, we
find
\begin{eqnarray}
{W} & = & \alpha \tau^{2/5} + \overline{\alpha}\tau^{4/5} + O(\tau^{6/5}), \\
V & = & \beta \tau^{2/5} + \overline{\beta}\tau^{4/5} + O(\tau^{6/5}),
\label{A5}
\end{eqnarray}
where $\alpha= (25/6)^{1/5} (\Gamma^3/\Gamma')^{1/10}$, $\beta = [25\Gamma'^2 /(6
\Gamma)]^{1/5}$, $\Gamma, \Gamma' > 0$, and
\begin{eqnarray}
\overline{\alpha}&=&(5/28)\left(3 \beta^{-3} - 8 J_1 \alpha^{-3}\right),\\
\overline{\beta}&=& (5/28)\left(6 J_1 \alpha^{-3} - 11 \beta^{-3}\right).
\end{eqnarray}

(3) The second and third term in equation~(\ref{A1}) cancel when
$V=\Gamma/J_1+\beta \tau^{2/3}$, $W=\alpha \tau^{2/3}$. However, we find that
$\alpha^4=-9 J_1\beta/(2\Gamma)$ and $\alpha^2=9 \Gamma' J_1^2/(2\beta\Gamma^2)$.
As $J_1 > 0$, and $\Gamma, \Gamma'$ take the same sign,
no real $\alpha$, $\beta$ pair satisfies these equations.
This possibility
is ruled out on the level of the coefficients.

On the other hand, spreadout, when it occurs, is such that $W$ and $V$ are proportional
to $t$ at large times. Alternatively, just $W \sim t$ and $V_0 \rightarrow \const$

\section{Conclusions}

Only for strong confinement is the Q2D analysis an adequate approach. For moderate
confinement the 3D region of stability is richer from that of Q2D (see
figure~\ref{fig3a}). A large, new region of stability appears in the 3D treatment as
compared to the Q2D treatment. This is proof of the fact that our solitons are truly 3D.
We are more used to the effect of adding a new dimension simply shrinking or abolishing
the basin of stability of solitons or waves. For example, water waves are unstable with
respect to perturbations along their direction of propagation only when the depth exceeds
a critical value (Benjamin \& Feir 1967). When, however, two dimensional perturbations
are allowed, there will always be an unstable angle regardless of the depth (Hayes 1973).
Another example is that of 1D solitons of the Nonlinear Schr\"{o}dinger equation (NLS)
with constant coefficients. They are stable in 1D, but unstable in 2D or 3D. This is also
true for some NLS waves (Anderson \textit{et. al.} 1979; Infeld \& Rowlands 1980, 2000).
In the problem treated here this is not the case. For situations coresponding to much of
the quasi two dimensional stability chart adding a degree of freedom {\it stabilizes} the
soliton solution. The key to this dichotomy is the presence of a periodic modulation,
absent in the above mentioned classical examples. This can be illustrated by a simple
case involving oscillators. Take as the one dimensional version a forced oscillator
problem:
\begin{equation}
\ddot{x}+\omega_0^2x=y\cos(\omega_0t).
\end{equation}
If $y$ is fixed, the solution has a secular component $yt/(2\omega_0)\sin(\omega_0t)$,
and so the amplitude will grow as $t$. If however we allow a second degree of freedom,
such that $y$ also oscillates (for instance $\ddot{y}+\epsilon^2y=0$) the solution
stabilizes, unless $\epsilon = \pm 2\omega_0$. In general this can be the case when there
are periodic modulations. This fact, obvious in oscillator theory, is perhaps less well
known in the soliton context.

The main result of this paper is pointing out the possibility of creating fully 3D
breather solitons in a BEC confined by a 1D optical lattice potential, corresponding to
the upper region in figure~\ref{fig3a}. The stable patterns may feature a multi-cell
structure. Depending on the strength of the confinement we identified three different
cases: a) strong confinement; practically no evolution in the $z$ direction, thus Q2D can
be applied, b) moderate confinement; fully 3D soliton-train, but no interaction between
cells, c) weak confinement; solution in  form a set of weakly interacting fundamental
solitons. The scheme proposed here is based on a combination of FRM and a 1D optical
lattice, and could be implemented in an experiment relatively easily. This would open the
way to the creation of robust 3D solitons (breathers) in BECs.

%To create stable 3D solitons in an experiment, one needs an OL of
%strength $\varepsilon_f$ of at least $10
%E_{\mathrm{recoil}}$. We define $\varepsilon_{\mathrm{rel}}=2 \varepsilon_f /
%E_{\mathrm{recoil}}$. The modulation frequency $\Omega$ must be in the range
%$[\Omega_0,2\Omega_0]$, where $\Omega_0=\frac{2\pi^2 \hbar}{m \lambda^2}
%\sqrt{4\varepsilon_{\mathrm{rel}}- \sqrt{\varepsilon_{\mathrm{rel}}}}$ is the excitation frequency
%of the confining potential. The final value of the mean scattering length $\overline{a_0}$ and
%number of atoms $N$ must satisfy the condition  $\overline{a_0}<a_c=-n
%\lambda\varepsilon_{\mathrm{rel}}^{1/4}/(2 \sqrt{\pi}N)$. The scattering length modulations $a_1$
%should be a few times larger than $|\overline{a_0}|$. Finally the initial radial optical trap
%frequency $\omega_{\perp}$ should be smaller than $\omega_{\perp 0}=m \lambda^3
%(\frac{\overline{a_0}}{a_c}-1)(\Omega^2-\Omega_0^2)/(2 N a_1 \varepsilon_{\mathrm{rel}} \hbar
%\pi^{5/2})$.

\begin{acknowledgements}

M.M. acknowledges support of the KBN grant 2P03 B4325, M.T. was supported by the Polish
Ministry of Scientific Research and Information Technology under grant PBZ
MIN-008/P03/2003, and E.I. acknowledges support of grant 2P03B09722. The authors are
deeply indebted to Professor Boris Malomed for valuable discussions.

\end{acknowledgements}

\label{lastpage}

\end{document}